\documentclass{article}
\usepackage{cite}
\usepackage{graphicx}
\usepackage{dcolumn}

\begin{document}

\date{}
\title{On the application of the Adomian-based methods}
\author{Francisco M. Fern\'{a}ndez\thanks{%
fernande@quimica.unlp.edu.ar} \\
INIFTA, DQT, Sucursal 4, C.C. 16, \\
1900 La Plata, Argentina}
\maketitle

\begin{abstract}
We analyse a recent application of two Adomian-based approximate methods to
the Chen-Lee-Liu equation. We prove that the outcome of these approaches is
merely the Taylor expansion of the solution about $t=0$ and, consequently,
they are valid only for sufficiently small values of the time variable $t$.
\end{abstract}

In a recent paper Mohammed and Bakodah (MB)\cite{MB21} compared the
performance of the Adomian decomposition method (ADM) and improved Adomian
decomposition method (IADM) on the nonlinear partial differential equation
\begin{equation}
iu_{t}+au_{xx}+ib|u|^{2}u_{x}=0,  \label{eq:dif_eq_1}
\end{equation}
and concluded that the later approximate method is slightly more accurate.
In order to determine the accuracy of these algorithms the authors resorted
to the exact solution\cite{TZMUBB18}
\begin{equation}
u(x,t)=\sqrt{\gamma \pm \eta \,\mathrm{sech}[\lambda (x-\nu t)]}e^{i[\omega
t-kx+\theta (x-\nu t)]},  \label{eq:u(x,t)_exact}
\end{equation}
where $\gamma $, $\eta $ and $\lambda $ depend on the model parameters $a$
and $b$ as well as on some arbitrary constants $k$, $B$, $\nu $ and $\omega $%
\cite{MB21}. MB did not comment on the, apparently necessary, extra phase
function $\theta (s)$ of the travelling variable $s=x-\nu t$\cite{TZMUBB18}.

MB showed tables with absolute errors and figures for some real function
that they failed to indicate (probably $|u(x,t)|$ or $|u(x,t)|^{2}$). They
chose $a=5$ and $b=10$ for their calculations but omitted the selected
values of $k$, $B$, $\nu $ and $\omega $. Therefore, it is not possible to
reproduce their results obtained for $t=0.3$ and $t=0.5$. This fact does not
affect the aim of the present Comment that is to show that they did not
choose larger values of $t$ because neither the ADM or the IADM are expected
to yield acceptable results for sufficiently large values of the time
variable.

For simplicity, we consider the ADM in what follows. MB did not show the
application of the ADM to equation (\ref{eq:dif_eq_1}) but to the closely
related one
\begin{equation}
u_{t}+iau_{xx}-b|u|^{2}u_{x}=0.  \label{eq:dif_eq_2}
\end{equation}
Curiously, they did not comment on the connection between the solutions to
these two nonlinear partial differential equations. The approach is based on
an expansion of the solution as
\begin{equation}
u(x,t)=\sum_{j=0}^{\infty }u_{j}(x,t),\;u_{0}(x,t)=u(x,0),
\label{eq:u_expansion}
\end{equation}
from which it follows that $A=|u|^{2}u_{x}$ can be expanded as
\begin{equation}
A(x,t)=\sum_{j=0}^{\infty }A_{j}(x,t).  \label{eq:A_expansion}
\end{equation}
MB also forgot to say how many terms of the expansion (\ref{eq:u_expansion})
were considered in their calculations and, most importantly, to test the
rate of convergence of the algorithms. From MB's equation (3.6) one easily
shows that
\begin{equation}
A_{n}=\sum_{j=0}^{n}u_{n-j\,x}\sum_{k=0}^{j}u_{k}\bar{u}_{j-k},
\label{eq:A_n}
\end{equation}
where $\bar{u}_{j}$ is the complex conjugate of $u_{j}$. Curiously, this
simple and efficient closed-form expression has not been taken into account
neither in this paper or in earlier ones where the authors have been doing
this same calculation again and again\cite{MBBAZBMB19,MB20a,MB20b}. One can
easily verify that equation (\ref{eq:A_n}) already yields the expressions
given in one of those papers for $n=1,2,3,4$\cite{MB20a}.

The ADM corrections are given by the recurrence relation\cite{MB21}
\begin{equation}
u_{k+1}(x,t)=ai\int_{0}^{t}u_{k\,x}(x,t^{\prime })\,dt^{\prime
}-b\int_{0}^{t}A_{k}(x,t^{\prime })\,dt^{\prime },\;k\geq 0.
\label{eq:u_(k+1)_ADM_1}
\end{equation}
Since $u_{0\,x}(x,t)$ and $A_{0}(x,t)=|u(x,0)|^{2}$ are independent of $t$,
then $u_{1}(x,t)$ is of the form $u_{1}(x,t)=v_{1}(x)t$, where $%
v_{1}(x)=iau_{0\,x}(x,0)-b|u(x,0)|^{2}$. In the next step we have $%
A_{1}(x,t)=u_{1}(x,t)\bar{u}_{0}(x,t)u_{0\,x}(x,t)+u_{0}(x,t)\bar{u}%
_{1}(x,t)u_{0\,x}(x,t)+u_{0}(x,t)\bar{u}_{0}(x,t)u_{1\,x}(x,t)=B_{1}(x)t$ so
that $u_{2}(x,t)=v_{2}(x)t^{2}$ where $v_{2}(x)$ is straightforwardly given
by equation (\ref{eq:u_(k+1)_ADM_1}). Repeating this argument we conclude
that $u_{j}(x,t)=v_{j}(x)t^{j}$ and $A_{j}(x,t)=B_{j}(x)t^{j}$, $%
j=0,1,\ldots $. Therefore, the expansion (\ref{eq:u_expansion}) reads
\begin{equation}
u(x,t)=\sum_{j=0}^{\infty }v_{j}(x)t^{j},  \label{eq:u_t-expansion}
\end{equation}
and shows that the ADM simply produces the Taylor expansion of $u(x,t)$
about $t=0$. Consequently, the ADM approximation to the solution $u(x,t)$
will be valid only for sufficiently small values of $t$. MB's choices $t=0.3$
and $t=0.5$ do not reveal the shortcomings of the method. In earlier
applications of similar approaches to the same type of equations the authors
also chose values of the time variable in the range $0.1\leq t\leq 0.5$\cite
{MBBAZBMB19,MB20a,MB20b}. It is worth mentioning that the functions $%
v_{j}(x) $ can be obtained recursively from
\begin{equation}
v_{k+1}(x)=\frac{1}{k+1}\left[ aiv_{k\,x}(x)-bB_{k}(x)\right] ,
\label{eq:v_(k+1)_ADM_2}
\end{equation}
that appears to be somewhat simpler than equation (\ref{eq:u_(k+1)_ADM_1}).

The IAM consists of writing $u(x,t)=u_{1}(x,t)+iu_{2}(x,t)$, $u_{1}$ and $%
u_{2}$ real, and applying the ADM to the real and imaginary parts of
equation (\ref{eq:dif_eq_2}) separately. A simple analysis, similar to the
one above, shows that in this case one obtains the $t$-power series for $%
u_{1}(x,t)$ and $u_{2}(x,t)$ and, consequently, for $u(x,t)$. Therefore, the
IADM exhibits the same drawback as the ADM.

Other authors have unknowingly derived Taylor series by means of unsuitable
implementations of approximate methods\cite{F09a,F09b,F10,F14,F20}.

\end{document}